\documentstyle[twocolumn,aps,prl,epsf]{revtex}
\def\Journal#1#2#3#4{{#1} {\bf #2}, #3 (#4)}

\def\PRL{ Phys. Rev. Lett.}
\def\PRD{{ Phys. Rev.} D}
\def\MN{ Mon. Not. R. Astron. Soc.}
\def\ApJ{ ApJ.}
\def\RMP{ Rev. Mod. Phys.}
\def\N{ Nature}

\def\be{\begin{equation}}
\def\ee{\end{equation}}
\def\bea{\begin{eqnarray}}
\def\eea{\end{eqnarray}}

\def\sst{\scriptscriptstyle}
\def\tr{\mathop{\rm Tr}\nolimits}

\begin{document}
\draft

%%%%%%%%%%%%%%%%%%%%%%%%%%%%%%%%%%%%%%%
%                Title                %
%%%%%%%%%%%%%%%%%%%%%%%%%%%%%%%%%%%%%%%
\twocolumn[\hsize\textwidth\columnwidth\hsize\csname
@twocolumnfalse\endcsname
\title{ Statistical Mechanics of
Self--Gravitating System : Cluster Expansion Method
}

\author{
Osamu Iguchi$^{a}$,
Tomomi Kurokawa$^{b}$,
Masahiro Morikawa$^{c}$,
}

\address{
Department of Physics,
Ochanomizu University, 1--1 Otuka,2 Bunkyo--ku,
Tokyo 112, JAPAN}

\author{
Akika Nakamichi$^{d}$,
}

\address{
Gunma Astronomical Observatory, 1--18--7
Ohtomo,  Maebashi,  Gunma 371, JAPAN}

\author{
Yasuhide Sota$^{e}$,
Takayuki Tatekawa$^{f}$, and 
Kei--ichi Maeda$^{g}$
}

\address{
Department of Physics, Waseda
University, 3--4--1 Okubo, Shinjuku--ku, Tokyo 169,
JAPAN}
\date{\today}
\preprint{}
\maketitle
\begin{abstract}
We study statistical mechanics of 
the self--gravitating system 
applying the cluster expansion method developed 
in solid state physics.
By summing infinite series of diagrams, 
we derive a complex free energy 
whose imaginary part is related to 
the relaxation time of the system.
Summation of another series yields 
two--point correlation function 
whose correlation length is essentially given 
by the Jeans wavelength of the system.
\end{abstract}

]

\narrowtext

%%%%%%%%%%%%%%%%%%%%%%%%%%%%%%%%%%%%%%%
%            Body of paper            %
%%%%%%%%%%%%%%%%%%%%%%%%%%%%%%%%%%%%%%%

%%%%%%%%%%%%%%%%%%%%%%%%%%%%%%%%%%%%%%%
%             Introduction            %
%%%%%%%%%%%%%%%%%%%%%%%%%%%%%%%%%%%%%%%
\paragraph{Introduction}

Variety of structures found in the present Universe are
thought to be seeded by quantum fluctuations 
in the very early Universe and 
evolved through the gravitational interaction 
in the expanding Universe aided by yet unknown Dark Matter.
Though there would be several aspects 
in the study of the formation of these rich structures,
we believe the self--gravity is 
one of the most intrinsic aspects
for the structure formation.

Therefore, in this paper, 
we would like to study fundamental
properties of the self--gravitating system (SGS) 
from a view point of statistical mechanics.
We specify the SGS as a mass distribution 
which consists of discrete mass points mutually interacting 
through the force of Newtonian gravity.
The contact collisions of the ingredients will be neglected.

The SGS is well known to be unstable and eventually collapses. 
Then it is widely believed that 
there is no thermal equilibrium in SGS.
However relevant structures would be observed 
in the intermediate stage of its evolution 
well before the collapse.
Therefore we may have a room to construct 
statistical mechanics of SGS. 
Actually there have been many studies dealing with 
statistical mechanics of SGS\cite{saslaw85}.
For example, 
an approach by de Vega {\it et al.}\cite{vega-C}
considering canonical/grand canonical ensemble of SGS 
in order to explain the scaling relation 
in the interstellar medium seems interesting.
The instability of SGS requires 
the inevitable introduction of cutoff in their work.
We would like to circumvent 
this situation allowing complex free energy, 
whose imaginary part provides 
dynamical information of the system.

We shall construct statistical mechanics of SGS 
after the model of classical electron--gas system 
on uniform ion background. 
It has the same square--inverse law of force
but with opposite signature to SGS.
Contrary to the ordinary belief, the correspondence,
i.e. the replacement $ e^2 \rightarrow -Gm^2$, 
where $e$ is the charge of electron and 
$G$ is the Newton's constant, 
%with the phase rotation from $0$ to $\pi$, 
is quite suggestive
if we identify the temperature $T$ 
as the velocity dispersion of SGS.
For example, 
the Debye wavelength $\lambda_{D}$ corresponds to 
the Jeans wavelength $\lambda_{J}$, 
and the plasma frequency is related to 
the inverse of the free fall time, 
$\omega_p^2 := -4\pi G m n =-1/\tau_{ff}^2$.
We believe that these correspondences are not merely appearance.
Actually, we can calculate the free energy of SGS 
in the same way as the classical electron gas system 
by applying the cluster expansion method
\cite{wortis83,mayer77,abe91}.
Moreover, $n$--point correlation functions are similarly obtained.

%%%%%%%%%%%%%%%%%%%%%%%%%%%%%%%%%%%%%%%
%     Cluster expansion method        %
%%%%%%%%%%%%%%%%%%%%%%%%%%%%%%%%%%%%%%%
\paragraph{Cluster expansion method}

We consider a non--relativistic gas system of
$N$ particles with the same mass $m$ 
in a box of size $L$.
The gas is in thermal equilibrium at $T$, 
where the temperature $T$ is identified with the velocity
dispersion $\langle {v^2}\rangle$ of the system, i.e.
$k_B T=m\langle {v^2} \rangle/3$,
where $k_B$ is the Boltzmann's constant.

We work in the canonical ensemble with 
the Hamiltonian of the system:
\be
H = \sum\limits_{i=1}^N {{ \vec p_i^{~2}} \over {2m}} 
    +\sum\limits_{1\le i<j \le N} {\phi _{ij}},
\label{Hamiltonian}
\ee
where
$\phi _{ij}$ is a potential through which each particles interact.
The partition function for fixed number of particles $N$,
volume $V=L^3$, and temperature $T$, is given by
\bea
Z(u) 
 &=& \tr\exp \left[-\beta H 
     + \sum\limits_{i=1}^{N} u(\vec x_i)\right] \cr
 &=& {{(2\pi m/\beta )^{3N/2} V^N} 
     \over {N! h^{3N}}}  e^{W(u)}, \\
\label{def-Z}
e^{W(u)} 
 &=& \int_{V^N}{{d^N\vec x} \over { V^N }} \exp
     \left[{\sum\limits_{i=1}^N {u(\vec x_i)}
     +\sum\limits_{1\le i<j \le N} {\varphi _{ij}}} \right],
\label{def-W}
\eea
where $\beta = (k_B T)^{-1}$, $h$ is the Planck's constant, 
and $\varphi_{ij} := -\beta \phi_{ij}$.
We have introduced a source function $u(\vec x_i)$ 
for convenience of calculation and integrated over the momenta.
By using functional derivative with respect to
the source $u(\vec x_i)$,
$k$--point correlation functions are given by
\be
G(\vec x_1,\cdots,\vec x_k) =
 \lim_{u \to 0}{{\delta ^k\ln Z(u)}
 \over {\delta u(\vec x_1)\cdots
 \delta u(\vec x_k)}}.
\label{kpf}
\ee

For the calculation of thermodynamic quantities,
we introduce the logarithm of the average of 
the configurational sum setting $u\to 0$:
\bea
 W_0 := W(0) &=&
 \ln\left\langle\exp \left(
 \sum\limits_{1\le i<j \le N} 
 {\varphi _{ij}} \right)\right\rangle,
 \label{def-W_0}
\eea
where 
$\left\langle \cdots \right\rangle 
= V^{-N}\int\! {\cdots d^N\vec x}$.

The cumulant expansion of $W_0$ is possible:
\bea
W_0 &=& \left\langle {\exp \left( 
        {\sum\limits_{1\le i<j \le N}{\varphi_{ij}}}
        \right)-1} \right\rangle_C \cr
&& \cr
    &=& \sum\limits_{ \sst {\nu_1\cdots \nu_M}\atopwithdelims()
        {\mbox{\tiny except all }\sst \nu_{i} =0}\hfill}
        {{{\left\langle {\varphi _1^{\nu _1}\varphi_2^{\nu_2}\cdots
        \varphi _M^{\nu _M}} \right\rangle _C} 
        \over {\nu _1! \cdots \nu_M! }}},
\label{w-cumulant}
\eea
where $\left\langle \cdots \right\rangle_C$ is 
a cumulant and $1,2,\cdots, M=N(N-1)/2$ are 
all the possible pairs of $N$--particles.

It is convenient to use graphical representations
in order to calculate $W_0$.
An each term in Eq.(\ref{w-cumulant}) corresponds to 
a graph which consists of vertices and lines.
Each line terminates at two distinct vertices.
A vertex $\vec x_i$ simply represents 
an integration position and 
a line connecting $\vec x_i $ and 
$\vec x_j $ represents an interaction $\varphi_{ij}$.
All the elements in a graph are multiplied with each other 
and integrated over all vertices.
This is the {\it cluster expansion method}. 
All the graph which contributes to $\ln Z$ is connected and 
does not include articulation points
which divide the graph into plural pieces, 
because of the character of the cumulant and the
translational invariance of the system.

By expanding with respect to the number density $n = N/V$,
$W_0$ is reduced to the following form\cite{wortis83,mayer77,abe91}.
\bea
W_0 &=& 
 \sum\limits_{k=1}^\infty {{{n^{k+1}} \over
 {(k+1)! }}}\sum\limits_{\mbox{\tiny connected}
 \hfill\atopwithdelims()
 \mbox{\tiny irreducible}\hfill} \times \cr
&& \cr
&& {\int_{V^{k+1}}{\prod\limits_{1\le i<j\le k+1}
 {\left[{{{(\varphi_{ij})^{\nu_{ij}}} \over
 {\nu_{ij}! }}} \right]d\vec x_1...d\vec x_{k+1}}}},
\label{w-expansion}
\eea
where $\nu_{ij}$ is the number of interaction lines
between the vertices $i$ and $j$.

%%%%%%%%%%%%%%%%%%%%%%%%%%%%%%%%%%%%%%%
%        Application to the SGS       %
%%%%%%%%%%%%%%%%%%%%%%%%%%%%%%%%%%%%%%%
\paragraph{Application to the self--gravitating system}

Let us consider a non--relativistic self--gravitating system 
where $\phi_{ij} = -Gm^2/r_{ij}$ with
$r_{ij} = \left| {\vec x_i-\vec x_j} \right|$.

\subparagraph{Free energy}

\begin{figure}[htbp]
 \begin{center}
   \leavevmode
   \epsfysize=5.0cm
   \epsfbox{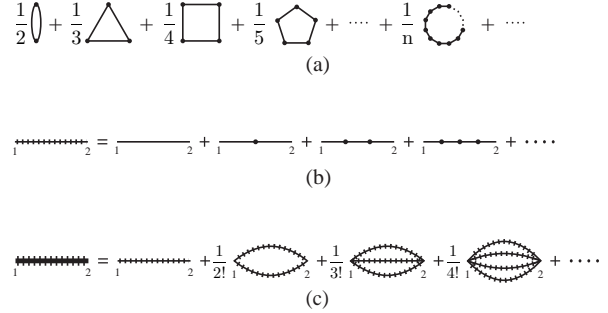}
\caption{
A series of cluster expansion:
(a) ring graphs.
(b) chain graphs.
(c) multi--lines of the chain graphs.
A symmetric factor is written 
in the left side of each graph.
}
 \label{cluster}
 \end{center}
\end{figure}

We shall include the higher orders in $n$ for calculating $W_0$.
Here we choose a series of graphs which is the lowest order of
$\varphi$ in each set of graphs containing $k$ internal vertices.
This is equivalent to a sum of all graph which 
has the topology of a ring 
(ring approximation shown Fig.\ref{cluster}--(a)). 
This approximation will not be valid for short distances 
where $\varphi$ grows without bound.

%%%%%%%%%%%%%%%%%%%%% fig %%%%%%%%%%%%%%%%%%%%%%%%%%%
%%%%%%%%%%%%%%%%%%% cluster %%%%%%%%%%%%%%%%%%%%%%%%%
%%%%%%%%%%%%%%%%%%%%% fig %%%%%%%%%%%%%%%%%%%%%%%%%%

A ring graph which contains $k$ vertecies in
Eq.(\ref{w-expansion}) corresponds to
\be
\int_{V^{k}}{\varphi _{12}\varphi _{23}\cdots
 \varphi_{k1}d\vec x_1...d\vec x_{k}}
   = \int_{|\vec q|>2\pi/L} {{{d\vec q} \over {(2\pi )^3}}
     \left[{\widetilde \varphi (\vec q)} \right]^k},
\label{ring}
\ee
where $\widetilde{\varphi} (\vec q) = 
{4\pi \beta Gm^2}/ {\vec q^{~2}}$ 
is the Fourier transform of $\varphi(\vec x)$.
$(k-1)! /2$ terms of this kind for each $k$ yield
\bea
W_0 = &&{V \over 2}\int_{|\vec q|>2\pi/L} {{{d\vec q} \over
         {(2\pi )^3}}}\sum\limits_{k=2}^\infty
         {{{\left[n\widetilde \varphi (\vec q)\right]^k} \over k}} \cr
    = && {V\over {4\pi ^2}}\int_{2\pi/L}^\infty
         {dq\cdot q^2}\left[ {-{{\kappa ^2} \over
         {q^2}}-\ln \left( 
         {1-{{\kappa ^2} \over {q^2}}} \right)}\right]\cr
    = && {1 \over 12\pi^2 }\Bigl[2\pi \kappa^2 L^2
         +\kappa^3L^3\ln \Bigl|
         {\kappa L+2\pi\over \kappa L-2\pi }\Bigr| \cr 
&&\quad \quad
         +8\pi^3 \ln\Bigl|1-{\kappa^2 L^2 \over 4\pi^2}\Bigr| \cr 
&&\quad \quad
         +i\pi\left(\kappa^3L^3  +8\pi^3 \right)
         \theta\left(\kappa L-2\pi\right)\Bigr],
\label{w-L}
\eea
where $\kappa ^2 := 4\pi\beta Gm^2 n=12\pi^2/\lambda_{J}^2$ 
with $\lambda_{J}=\sqrt{\pi \langle v^2 \rangle /(Gmn)}$
being the Jeans wavelength and $\theta$ is the Heaviside function. 
For the integration in Eq.(\ref{w-L}), 
we analytically continue $e^2$ to $-Gm^2$ from 
the above complex $e^2$ plane by the reason under Eq.(\ref{w}).

We notice that the imaginary part appears for
$ L > 2\pi /\kappa =\lambda_{J} /\sqrt{3}$.
This originates from the negative argument 
in the logarithm for small momentum $q$.
The appearance of it apparently indicates the Jeans instability.
The thermodynamic limit $N \to \infty, V \to \infty$ with fixed
number density $n$, yields the free energy:
\bea
F &=& -\beta^{-1}\ln Z(0) \cr
  &=& -{N\over\beta}\left[{3 \over 2}\ln \left( {{2\pi m}
    \over {\beta h^2}} \right)+1
    -\ln n-{{i\kappa ^3} \over {12\pi n}}\right].
\label{w}
\eea
The above choice of analytical continuation guarantees 
the positivity of the imaginary part of $F$: 
The system is truly dissipative rather than anti--dissipative.

In general, the imaginary part of a free energy is directly
related with the decay strength of the system 
$\Gamma$\cite{affleck81}:
$\Gamma=(\omega \beta/\pi) $Im$F $,
where $- \omega^2$ is the negative eigenvalue 
at the saddle point which divides 
a metastable region from a stable region.
If we identify $\omega$ as the inverse of the free fall time,
i.e. $\omega = \tau_{ff}^{-1} = \sqrt{4\pi G mn}$, 
the decay strength $\Gamma$ becomes
\be
\Gamma = { {4 \sqrt{3} N G^2m^2n}
         \over {\left\langle {v^{2}} \right\rangle ^{3/2}}},
\label{decay}
\ee
which is essentially the inverse of 
the binary relaxation time\cite{saslaw85,chan43},
\be
\tau _{bc} = 
 {{\left\langle {v^{2}}\right\rangle ^{3/2}}
 \over 32\pi G^2m^2n\ln (N/2)},
\ee
except replacing $\ln N$ with $N$.

Contrary to the extensive variables 
in the conventional thermodynamics,
the imaginary part of $F$ apparently is superextensive
since ${\rm Im} F \propto n^{3/2}$ but not $\propto n$.
Moreover the imaginary part is related to the fluctuation 
of the system through the fluctuation--dissipation theorem.
In this context, it seems interesting to notice that
the number $3$ in the above $n^{3/2}$ 
comes from the spatial dimensionality
and $2$ from the inverse--square law of the gravitational force.
This reminds us of the Holtsmark distribution of the gravitational
force acting in the uniform self--gravitating system \cite{chan43} 
or the stable distribution of index $3/2$ \cite{feller66}.

%%%%%%%%%%%%%%%%%%%%%%%%%%%%%%%%%%%%%%%
%   Two-point correlation function    %
%%%%%%%%%%%%%%%%%%%%%%%%%%%%%%%%%%%%%%%
\subparagraph{Two--point correlation function}

Let us now turn our attention to the correlation functions.
One--point correlation function is simply a number: $G_1(\vec x) = N$.
The normalized two--point correlation function
which is usually used in astrophysics is
\be
\xi (r) = 
 G(\vec x_1,\vec x_2)/\left[G(\vec x_1)G(\vec x_2)\right],
\ee
where $r=|\vec x_1-\vec x_2|$.
A similar cluster expansion for $ G(\vec x_1,\vec x_2)$ would be
\bea
G(\vec x_1,\vec x_2) &=&  
 N(N-1) \sum\limits_{k=0}^\infty  {{{n^{k}}
 \over {k! }}}\sum\limits_{\mbox{\tiny connected}
 } \times \cr
&& \cr
&&
 \int_{V^{k}}{\prod\limits_{1\le i<j\le k+2}
 {\left[{{{(\varphi _{ij})^{\nu _{ij}}} \over
 {\nu _{ij}! }}} \right]d\vec x_3...d\vec x_{k+2}}}.
\label{G-expansion}
\eea
We shall include higher orders in 
$n$ as previous calculation.
Therefore, among each set of graphs 
which contain $k$ internal vertices 
(excluding both ${\vec x_1}$ and ${\vec x_2}$), 
we choose the lowest order skeleton graph in $\varphi_{ij}$.
This is equivalent to a sum of all the graph which 
has the topology of a chain shown in  Fig.\ref{cluster}--(b).
%--- chain approximation
\bea
\overline \varphi _{12}
 &=& \varphi _{12}+\int_V {\varphi _{13}\varphi _{32} d\vec x_3} \cr
 &&  \quad\quad\!\! +\int_{V^2} {\varphi_{13}\varphi _{34}\varphi _{42} 
     d\vec x_3 d\vec x_4 +\cdots } \cr
 &=& \int\! {{d\vec q} \over {(2\pi )^3}}{{\widetilde \varphi (\vec q)}
     \over {1-n\widetilde \varphi (\vec q)}}
     \exp\left[i\vec q(\vec x_1 - \vec x_2)\right] \cr
 &=& {{2\beta Gm^2} \over {\pi r}}\int_0^\infty
     {d\lambda {{\lambda \sin \lambda}\over {\lambda ^2-\kappa ^2 r^2}}} \cr
 &=& \beta Gm^2{\cos  (\kappa r) \over r}.
\label{chain}
\eea
Since we have analytically continued %$e^2$ to $-Gm^2$ 
from the above complex $e^2$ plane in Eq.(\ref{w-L}), 
we should choose the pole at $- \kappa r$ for the integration.
%--- multi-line of chain
Summing over multi--lines of chain graphs 
shown in Fig.\ref{cluster}--(c),
we obtain the following function:
\bea
 \widehat \varphi _{12}
 &=& \overline \varphi _{12}
     +{1 \over {2! }}(\overline \varphi_{12})^2
     +{1 \over {3! }}(\overline \varphi _{12})^3
     +{1 \over {4! }}(\overline  \varphi _{12})^4+\cdots \cr
 &=& \exp (\overline \varphi _{12})-1.
\label{LF}
\eea
Moreover, we should consider 
mass renormalization for the external vertex.
In principle, this is given by sum of all the graph 
which consists of a cluster around 
the external vertex $\vec x_1$.
Since infinite summation of this class of graphs at present 
is technically impossible, 
we phenomenologically introduce an effective mass.
Since the gravitational attractive force balances with 
the stirring force arising from the velocity dispersion 
at the length scale $\kappa^{-1}$, 
the mass inside of this scale is thought to behave collectively.
Thus we estimate the effective mass $m^* = (4\pi /3)(\kappa ^{-1})^3nm$, 
which should be replaced with the mass $m$ at
$\vec x_1$ or at $\vec x_2$ but not both.
Strictly speaking, 
$m^*$ is a parameter in the present phenomenological argument.
By replacing $m$ at $\vec x_1$ in Eq.(\ref{chain}) with $m^*$ at
$\vec x_1$, the two--point correlation function becomes
$G(\vec x_1,\vec x_2) = N^2\left. {\widehat \varphi _{12}}
\right|_{m^{}_{\mbox{\tiny at} \vec x_1}
\to m^*_{\mbox{\tiny at} \vec x_1}}$.
Finally we obtain the normalized two--point correlation function:
\be
\xi (r)
  = \exp \left[{\cos (\kappa r) \over {3\kappa r}}\right]-1.
\label{2PF}
\ee
The scaling property is manifest:
Small scale SGS with large $\kappa$ has the same correlation function 
as that of large SGS with small $\kappa$.

This function $\xi(r)$ has interesting features.
As a function of $s=\kappa r$, 
it has a unique inflection point at $s=0. 440$ 
when plotted in the Log--Log graph.
The slope at $s=0.44$ is $-1.667$ 
and the magnitude there is $0.986$.
Therefore, 
the inflection point 
$r_c\sim 0.44 \kappa^{-1}\sim 0.04 \lambda_{J}$ 
is regarded as the correlation length of SGS.

%%%%%%%%%%%%%%%%%%%%%% fig %%%%%%%%%%%%%%%%%%%%%%%%%%
%%%%%%%%%%%%%%%%%%%%%% 2PF %%%%%%%%%%%%%%%%%%%%%%%%%%
\begin{figure}[htbp]
 \begin{center}
   \leavevmode
   \epsfysize=7.0cm
   \epsfbox{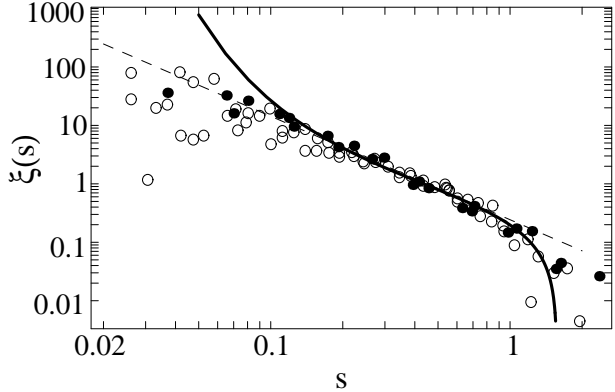}
\caption{
A Log--Log graph of two--pint correlation function: 
$\xi (s)=\exp\left[\cos (s)/(3s)\right]-1$ (solid line),
$\left(s/0.45\right)^{-1.77}$ (broken line),
observational data of 
galaxies~\protect\cite{davis86} (open circle),
and observational data of clusters of 
galaxies~\protect\cite{dalton94} (filled circle).
These correlation lengths of the observational data 
are rescaled. 
} 
 \label{fig-2PF} 
 \end{center}
\end{figure} 
%%%%%%%%%%%%%%%%%%%%% fig %%%%%%%%%%%%%%%%%%%%%%%%%%

The above two--point correlation function is shown in Fig.\ref{fig-2PF}
with typical observational results\cite{davis86,dalton94} on top of it.
The observation does not exclude our correlation function except
small scale region where the interaction $\varphi$ exceeds $1$ 
and the approximation we used is obviously invalid.

{}From the observational data\cite{davis86,dalton94}, 
the correlation lengths of galaxies and of clusters of galaxies 
directly read off as 
$r_{\sst c}^{\sst galaxy} \sim 6.2 h^{-1}{\rm Mpc}$ and 
$r_{\sst c}^{\sst cluster} \sim 15 h^{-1}{\rm Mpc}$ respectively.
Since $r_c\sim 0.44 \kappa^{-1}$ 
for the two--point correlation function we obtained,
we rescaled the observational data: $r \to s=\kappa r$
with $\kappa=1/14 ~h{\rm Mpc}^{-1}$ for galaxies and 
$\kappa=1/34 ~h{\rm Mpc}^{-1}$ for clusters of galaxies,
where $h = H_0/\left(100 {\rm km s}^{-1}\mbox{Mpc}^{-1}\right)$
with $H_0$ being the Hubble constant at the present time.
These values of $\kappa$ correspond to
$\lambda_{\sst J}^{\sst galaxy} \sim 152h^{-1}{\rm Mpc}$ and 
$\lambda_{\sst J}^{\sst cluster} \sim 370h^{-1}{\rm Mpc}$ 
from the relation given immediately after Eq.(\ref{w-L}).
On  the other hand, 
typical Jeans lengths calculating from the standard observations 
for galaxies and for clusters of galaxies are respectively 
\bea
\lambda_{\sst J}^{\sst galaxy} =
    123 {\rm Mpc} \Biggl[ &&
    \left({\left\langle {v^2} \right\rangle 
    \over 500 {\rm kms}^{-1}}\right)
    {\left({2\times 10^{44} {\rm g} 
    \over m}\right)} \times \cr
 && {\left({3.7\times 10^{-2} h^3{\rm Mpc}^{-3} 
    \over n}\right)}\Biggr]^{-1/2}, \\
    \lambda_{\sst J}^{\sst cluster} =
    450 {\rm Mpc} \Biggl[ &&
    \left({\left\langle {v^2} \right\rangle 
    \over 1000 {\rm kms}^{-1}}\right)
    {\left({2\times 10^{47} {\rm g} 
    \over m}\right)} \times \cr
 && {\left({1.1\times 10^{-5} h^3{\rm Mpc}^{-3} 
    \over n}\right)}\Biggr]^{-1/2}.
\eea
These values of the Jeans lengths
are not so far from the above values 
through two--point correlation function we obtained.

%%%%%%%%%%%%%%%%%%%%%%%%%%%%%%%%%%%%%%%
%         Summary and outlook         %
%%%%%%%%%%%%%%%%%%%%%%%%%%%%%%%%%%%%%%%
\paragraph{ summary and outlook }

%--- summary
Studying the canonical ensemble of 
the self--gravitating system (SGS), 
we obtained the complex free energy 
by summing an infinite series of graph 
in the cluster expansion method for SGS, Eq.(\ref{w}).
The imaginary part of the free energy yields 
the decay strength of SGS, Eq.(\ref{decay}).
Similar summation of an infinite series of graph yields 
the universal two--point correlation function Eq.(\ref{2PF}) 
which scales essentially with the Jeans wavelength.
The correlation length is linearly proportional to 
the mean separation of ingredients.

%--- outlook
We hope further development will be reported soon, 
including
(a) the higher--point correlation functions,
(b) much profound calculation on the complex free energy, 
(c) systematic argument on the mass renormalization $m^*$,
(d) the observational tests of our arguments, and 
(e) the effects of the cosmic expansion and of the Dark matter.

%%%%%%%%%%%%%%%%%%%%%%%%%%%%%%%%%%%%%%%
%             References              %
%%%%%%%%%%%%%%%%%%%%%%%%%%%%%%%%%%%%%%%

\end{document}